\documentclass[reprint,
superscriptaddress,
showpacs,
preprintnumbers,
nofootinbib,
nobibnotes,
amsmath,
amssymb, 
aps,
prd,
floatfix
]{revtex4-1}

\usepackage[utf8]{inputenc}
\usepackage[normalem]{ulem}
\usepackage{graphicx}
\usepackage{dcolumn}
\usepackage{bm}
\usepackage[colorlinks=true,allcolors=purple]{hyperref}
\usepackage{url}
\usepackage{enumerate}

\usepackage{slashed,multirow,relsize,soul,feynmp-auto,tikz}
\usepackage{color}
\usepackage{mathrsfs} 
\usepackage{amsmath}
\usepackage{cancel}

\usepackage{bbold}
 \usepackage{mathrsfs}
\usepackage{braket}
\usepackage{physics}
\usepackage{multirow}

\usepackage{fontawesome} 
\definecolor{blue-violet}{rgb}{0.33, 0.17, 0.89}
\newcommand{\refeq}[1]{Eq.~(\ref{#1})}
\newcommand{\refeqs}[2]{Eqs.~(\ref{#1})~and~(\ref{#2})}

\newcommand{\reffig}[1]{Fig.~\ref{#1}}
\newcommand{\reffigs}[2]{Figs.~\ref{#1}~and~\ref{#2}}
\newcommand{\refsec}[1]{Section~\ref{#1}}
\newcommand{\refapp}[1]{Appendix~\ref{#1}}
\newcommand{\reftab}[1]{Table~\ref{#1}}
\newcommand{\refref}[1]{Ref.~\cite{#1}}

\newcommand{\mbeq}{\overset{!}{=}}
\newcommand{\hdark}{{h_d}}
\newcommand{\Adark}{{\gamma_d}}
\newcommand{\nudark}{{N}}

\newcounter{CommentCount}
\definecolor{MH}{rgb}{0.0,0.6,9}
\definecolor{palatinate}{rgb}{0.494, 0.192, 0.482}

\renewcommand{\phi}{\varphi}
\newcommand{\orcid}[1]{\href{https://orcid.org/#1}{#1}}

\begin{document}

\preprint{\hfill FTPI-MINN-20-34, UMN-TH-4003/20}

\title{Novel multi-lepton signatures of dark sectors in light meson decays}

\author{Matheus Hostert}
\email{mhostert@perimeterinstitute.ca}
\thanks{\orcid{0000-0002-9584-8877}}
\affiliation{School of Physics and Astronomy, University of Minnesota, Minneapolis, MN 55455, USA}
\affiliation{William I. Fine Theoretical Physics Institute, School of Physics and Astronomy, University of
Minnesota, Minneapolis, MN 55455, USA}
\affiliation{Perimeter Institute for Theoretical Physics, Waterloo, ON N2J 2W9, Canada}
\author{Maxim Pospelov}
\affiliation{School of Physics and Astronomy, University of Minnesota, Minneapolis, MN 55455, USA}
\affiliation{William I. Fine Theoretical Physics Institute, School of Physics and Astronomy, University of
Minnesota, Minneapolis, MN 55455, USA}

\date{\today}

\begin{abstract}
We point out kaon decays to multiple charged leptons as a novel probe of light dark particles $X$. Previously  neglected channels, such as $K^+\to \pi^+ \left(XX \to 2(e^+e^-)\right)$, $K_L\to \pi^0 \left( XX \to 2(e^+e^-)\right)$, and $K_S\to  \left(XX \to  2(e^+e^-)\right)$ may have very large rates, exceeding not only the Standard Model expectations but also possible backgrounds, such as Dalitz decays of neutral pions. We apply this idea to dark sector models where the production of dark Higgses or heavy neutral leptons leads to final states with several visible dark photons. We also investigate a recently proposed model of an MeV-scale QCD axion, where the rates for kaon decays to multiple axion states are large due to the non-linear interactions of the axion with the light mesons. In addition, we point out new probes of this axion in pion decays, such as the single production of $a$ in $\pi^+ \to \nu \left((e^+)^* \to e^+ a\to e^+e^+e^-\right)$, double production in pion capture $\pi^-+(p \text{ or D})\to aa + (n\text{ or }nn) \to 2(e^+e^-) + (n\text{ or }nn)$, as well as $\pi^0 \to aaa\to 3(e^+e^-)$. The latter decay is fixed at $\mathcal{B}(\pi^0\to a a a) = 1.0 \times 10^{-3}$ for a $17$~MeV axion.
\end{abstract}

\maketitle

\section{Introduction} 

Low-energy extensions of the Standard Model (SM) typically operate with new states that are neutral under the SM group and have small couplings to SM particles. Collectively, such models are often referred to as ``dark sectors" (DS), as they might (or might not) be connected to the problem of dark matter. Considerable efforts of the last decade have identified leading dark sector models as well as the most sensitive experimental probes that deliver an ever-improving set of constraints on the parameter space of these models~\cite{Alexander:2016aln}.

One of the most stringent tests of the SM is delivered through the studies of kaon decays. Owing to their relative longevity, $K_L$ and $K^\pm$ decays can be studied away from the production point, which in many cases allows backgrounds to be reduced with respect to the signatures generated by the kaon decays themselves. Despite the seventy-year-long history of kaon studies, novel decay modes are being found and examined~\cite{Batley:2018hxd,NA62:2020fhy}, reaching down to $O(10^{-10})$ branching ratios (BR) in some cases. It is well known that such sensitivity can be used for searches of dark sector particles produced in kaon decays.

The models probed by the decays of flavored mesons, mostly $K$ and $B$, typically have specific properties. The single production of particles $X$ is most efficient when $X$ is {\em not} coupled to a conserved current, and consequently, the electroweak (EW) loop generating flavor-changing neutral currents (FCNC) can be enhanced by the masses of heavy particles, $t,W$, inside the loop. This way, strong constraints on DS particles $X$ can be obtained by studying $K\to \pi X$ and $B\to K^{(*)}X$ transitions, when $X$ is a Higgs-mixed scalar~\cite{Willey:1982mc,OConnell:2006rsp,Pospelov:2007mp,Batell:2009jf,Bezrukov:2009yw,Winkler:2018qyg} or an axion or axion-like particle~\cite{Anselm:1977jf,Freytsis:2009ct,Calibbi:2016hwq}, or dark vector with the mass mixing with $Z$-boson~\cite{Davoudiasl:2012ag,Dror:2018wfl}. In Refs.~\cite{Dror:2017nsg} it was shown that a generic DS vector coupled to non-conserved currents (non-conservation may come from the anomalous diagrams) is best constrained by flavor-changing decays. 

In comparison, pair production of dark states in the flavor decays is less explored. Original interest in $B\to K XX$ decays~\cite{Bird:2004ts,Bird:2006jd,Badin:2010uh} was driven by the minimal scalar dark matter model~\cite{Silveira:1985rk,McDonald:1993ex,Burgess:2000yq}, which has since been ruled out in this mass range. Other models also predict fully invisible neutral kaon decays~\cite{Gninenko:2014sxa,Gninenko:2015mea,Barducci:2018rlx}, such as $K_{S,L} \to X X$, which, despite its experimental challenges, could be searched for at NA64~\cite{NA64epps}. A recent increase of theoretical activity in studying dark sectors in kaon decays~\cite{Kitahara:2019lws,Egana-Ugrinovic:2019wzj,Dev:2019hho,Ziegler:2020ize,Gori:2020xvq,Hostert:2020gou,Altmannshofer:2020pjb} was instigated by a possible excess of $K_L\to\pi^0\nu\overline{\nu}$ signal events reported by the KOTO collaboration~\cite{kotoKAON19}. For instance, Ref.~\cite{Hostert:2020gou} identified a number of promising decay modes of $K_L$ to a pair of dark state particles $X_1X_2$, with subsequent decay of one or both of them to photon final states.

These studies have shown that kaon decays remain under-explored and that new sensitivity to DS can be derived. Thus far, most kaon physics studies are still driven by precision measurements of the SM decay modes. Yet, several modes that look ``hopeless" from the point of view of detecting a SM rate are still very promising for studying DS.
In this paper, we point to several new decay modes that deserve special attention, and that can be already analyzed using existing experimental data collected by {\em e.g.} NA48, NA62, KTeV, KOTO, KLOE, and their predecessors. In particular, we argue that the $K^+\to \pi^+ XX \to \pi^+ + 2(e^+e^-)$, $K_L\to \pi^0 XX \to \pi^0 + 2(e^+e^-)$, and $K_S\to  XX \to  2(e^+e^-)$ decay modes studied with precision on branching ratios at $\sim O(10^{-6})$ level and better can set new limits on DS models, or else discover new DS particles that have eluded detection thus far. We argue that mitigation of Dalitz background for these decay modes is easier than for {\em e.g.} $K\to \pi X\to \pi e^+e^-$. We also consider the possibility of final states with missing energy such as in decay chains involving heavy neutral leptons (HNLs) in the final state. 

For concreteness, we calculate kaon decay rates in two separate models. The first is based on higgsed and secluded $U(1)_d$ gauge-extensions of the SM, which can also admit couplings to additional singlet scalars or HNLs. We focus on Higgs portal and neutrino portal production of dark photons in multiplicities $\geq 2$. The subsequent prompt decays of the dark photons to electron-positron pairs leads to multi-lepton signatures. The second model we consider is that of a ``strongly coupled" QCD axion~\cite{Alves:2017avw,Alves:2020xhf} with $f_a \sim 1$\,GeV. This particle may evade constraints based on single axion production due to strong-interaction uncertainties in their direct mixing with $\pi^0$. We consider previously-neglected interactions of the axion with the pion that are quadratic or cubic in the axion field, allowing for independence from the value of axion-pion mixing angle predictions of kaon and pion decay rates. Perhaps the most striking prediction of GeV-scale axion model is a very large rate for decay $\pi^0\to 3a \to 3(e^+e^-)$ {\em exceeding} SM decay $\pi^0\to 2(e^+e^-)$ over a large fraction of parameter space. 

This paper is organized as follows. In Section II, without any specific model in mind, we introduce ``missing" (from the entire body of kaon literature) decay modes. In Section III, we point out new possible searches for dark sector particles. In Section IV, we consider new decay modes within a model for MeV-scale QCD axions with $f_a \sim {\rm GeV}$. We reach our conclusions in section V. 

\section{Novel multi-lepton kaon decays}
\label{sec:decays}
\renewcommand{\arraystretch}{1.4}
\begin{table*}[t]
    \centering
    \begin{tabular}{|l|c|c|cc|c|}
        \hline\hline
        Kaon decay & predicted BR in SM & measured BR & DS model & DS BR & additional resonances  \\\hline
\hline
        \multirow{2}{*}{$K_{S}\to 2(e^+e^-)$} & \multirow{2}{*}{$1.66\,(1.78) \times10^{-10}$ from \cite{DAmbrosio:2013qmd}} & \multirow{2}{*}{No search} & $U(1)_{d}+S$ & $2\times 10^{-10}$ & $-$ 
\\
        && & $a(17)$ axion & $\gtrsim 2.6\times 10^{-7}$ & $-$
\\
\hline\hline
     \multirow{2}{*}{$K_{L}\to 2(e^+e^-) $} &  \multirow{2}{*}{$3.65\times 10^{-8}$ from \cite{DAmbrosio:2013qmd}} &  $(3.72\pm0.29)\times 10^{-8}$ \cite{AlaviHarati:2001ab} & $U(1)_{d}+S$ & $2\times 10^{-8}$ & $-$
\\
    &&$(3.67\pm0.40)\times 10^{-8}$ \cite{Lai:2005kw}& $a(17)$ axion & $\gtrsim 7.2\times 10^{-10}$ & $-$
\\\hline
        \multirow{2}{*}{$K_{L}\to \pi^0 \, 2(e^+e^-)$} &  \multirow{2}{*}{N/A, expected $\mathcal{O}(10^{-10})$} &  \multirow{2}{*}{No search} & $U(1)_{d}$ & $2\times 10^{-7}$ & 
        $m_{4e}=m_{\hdark}$\\
        &&& $a(17)$ axion & $7\times 10^{-5}$ & $-$
\\
\hline
        $K_L\to \pi^0 \, 2(\nu e^+e^-)$ & N/A & No search & $U(1)_d +$HNL & $2\times 10^{-8}$ & $-$
\\\hline\hline
        \multirow{2}{*}{$K^+\to \pi^+ \, 2(e^+e^-)$} &  \multirow{2}{*}{N/A, expected $\mathcal{O}(10^{-10})$} &  \multirow{2}{*}{No search} & $U(1)_{d}$ & $5\times 10^{-8}$ & $m_{4e}=m_{\hdark}$\\
        &&& $a(17)$ axion & $1.7\times 10^{-5}$ & $-$
\\\hline
       $K^+ \to e^+ \nu \, 2(e^+e^-)$ & \multirow{2}{*}{N/A ($\ll 10^{-10}$)}& \multirow{2}{*}{No search} & \multirow{2}{*}{$U(1)_d +$HNL} & 
       $6 \times 10^{-8}$ & $m_{4e} = m_\hdark$, $m_{\nu 4e } = m_N$ 
\\
        $K^+ \to \mu^+ \nu \, 2(e^+e^-)$ & &  & & $9 \times 10^{-7}$ & $m_{4e} = m_\hdark$, $m_{\nu 4e } = m_N$ 
\\\hline
        $K^+\to \pi^+ \, 2(\nu e^+e^-)$ & N/A ($\ll 10^{-10}$)& No search & $U(1)_d +$HNL & $5\times10^{-9}$ & $m_{2\nu 4e} = m_\hdark$ \\\hline
\hline
        \end{tabular}
    \caption{Multi-lepton kaon decays channels and their current SM predictions and measurements, where available. We also show allowed branching ratios in new physics models (assuming $s_\theta \sim 5\times10^{-3}$, $m_S = 800$~MeV, $|U_{\mu 4}|^2 \sim 10^{-6}$, $|U_{\mu 4}|^2\sim10^{-7}$, and $m_N=150$~MeV for $U(1)_d$ models, and $m_a = 17$ MeV for the axion). In all new physics scenarios, $m_{ee}=m_{ee}^\prime = m_{X}$, where $X=\Adark$ in $U(1)_{d}$ models or $X=a$ for the MeV axion. Additional resonances that can be reconstructed on an event-by-event basis are also shown in the last column.}
    \label{tab:BRstable}
\end{table*}

Despite impressive progress on the search for light dark particles in pseudoscalar, $\pi^0,K$ and $B$ meson decays, much of the literature has been exclusively concerned with scenarios where mesons decay to a single new dark boson $X$ that is either invisible or decays electromagnetically into $\ell^+\ell^-$ or $\gamma\gamma$ pairs. On the other hand, much of the open parameter space of minimal dark sectors still allows for large BR of mesons to a pair or several visibly-decaying $X$ states. We will consider two classes of models where there is no need to confront large powers of small new physics couplings to emit several $X$ particles. Firstly, $X$ can be produced as a final state of some decay cascade in the dark sector, e.g. $K\to\pi(X^\prime\to X X)$. Alternatively, $X$ can be coupled quadratically or cubically to the SM ($\mathcal{L}\supset\mathcal{O}_{\rm SM} \times X^2 + \mathcal{O}_{\rm SM}^\prime\times X^3$) relatively strongly, as is the case for the couplings of a MeV-scale axion to light mesons.

Once produced, the multiple $X$ states may lead to several pairs of photon or lepton-antilepton final states, but for concreteness, we will focus on decays into final states with a pair of electron-positron final states, $2(e^+e^-)$. In this case, $X$ can be searched for by looking for an excess of events around a specific combination of di-lepton invariant masses: $m_{ee}=m_{ee}^\prime = m_{X}$ (a procedure known as ``bump hunt"). When the four leptons arise from a cascade in the dark sector (e.g. $X^\prime \to X X$), it will also be possible to find additional variables that reconstruct the masses of dark particles, as for example, $m_{4e} = m_{X^\prime}$. We note that this type of multi-lepton signature has been studied in the context of $e^+e^-$ colliders~\cite{Batell:2009yf,Essig:2009nc} and that experimental searches for $e^+e^-\to X(X^\prime\to X X) \to 6$~leptons have been performed at BaBar~\cite{Lees:2012ra} and Belle~\cite{TheBelle:2015mwa}, but the mass regime of $m_{X} < 100$~MeV with $m_{X^\prime} < 200$~MeV remains unexplored as of yet. This gap can be filled with kaon decays. 

First, we consider the production of pairs of $X$ alongside a pion. In this case, both neutral and charged kaons may offer an opportunity for discovery,
\begin{align}\label{eq:pi4e}
    K_{S,L} &\to \pi^0 \, X \, X \to \pi^0 \, 2(e^+e^-),
    \\
    K^+ &\to \pi^+ \, X \, X \to \pi^+ \, 2(e^+e^-).
\end{align}
No measurements or searches for the decays above exist, with or without bump hunts. The theoretical prediction within SM is also missing, although it is likely to be dominated by the $K\to\pi\gamma^*\gamma^*$ intermediate vertex. 
For comparison, its real photon counterparts appear at branching ratios of $\mathcal{B}(K\to\pi\gamma\gamma)\sim \mathcal{O}(10^{-6})$ and $\mathcal{B}(K\to\pi\gamma e^+e^-)\sim \mathcal{O}(10^{-8})$ in the SM. 
If so, the corresponding decay to two electron-positron pairs and a pion is unlikely to exceed $\mathcal{O}(10^{-10})$ benchmark, making it a perfect testing ground for new physics. 

The charged kaon channel $K^+\to\pi^+\gamma\gamma$ was measured at NA62~\cite{Ceccucci:2014oza}, $K^+\to\pi^+\gamma e^+e^-$ and $K_L\to\pi^0\gamma\gamma$ at NA48~\cite{Batley:2007uh,Lai:2002kf}, and $K_L\to\pi^0\gamma e^+e^-$ at KTEV~\cite{Abouzaid:2007xe}. Similarly, final states with one additional pion with respect to \refeq{eq:pi4e} have been studied at KTEV~\cite{Abouzaid:2008cd}, where a total of $3\times10^4$ pion double Dalitz decays ($\pi^0_{\rm DD} \to 2(e^+e^-)$) events were collected from $K_L\to\pi^0\pi^0\pi^0_{\rm DD}$ decays. The previous measurement, although not useful for the new physics we focus on, already shows that kaon experiments are reaching single event sensitivities to multi-lepton final states to branching ratios at the level of $10^{-10}$. Thus, in some distant future, even the SM rates for $K\to\pi 2(e^+e^-)$ may be observed.

\begin{figure*}[t]
    \centering
    \includegraphics[width=0.9\textwidth]{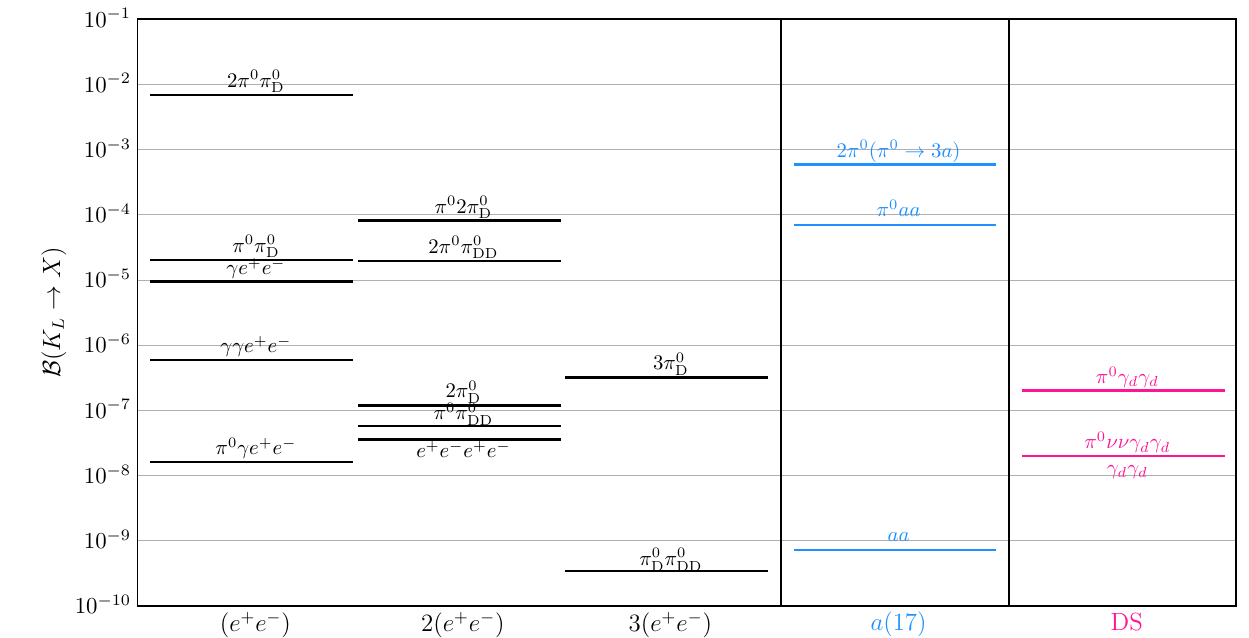}
   \\
\vspace{2ex}
    \includegraphics[width=0.9\textwidth]{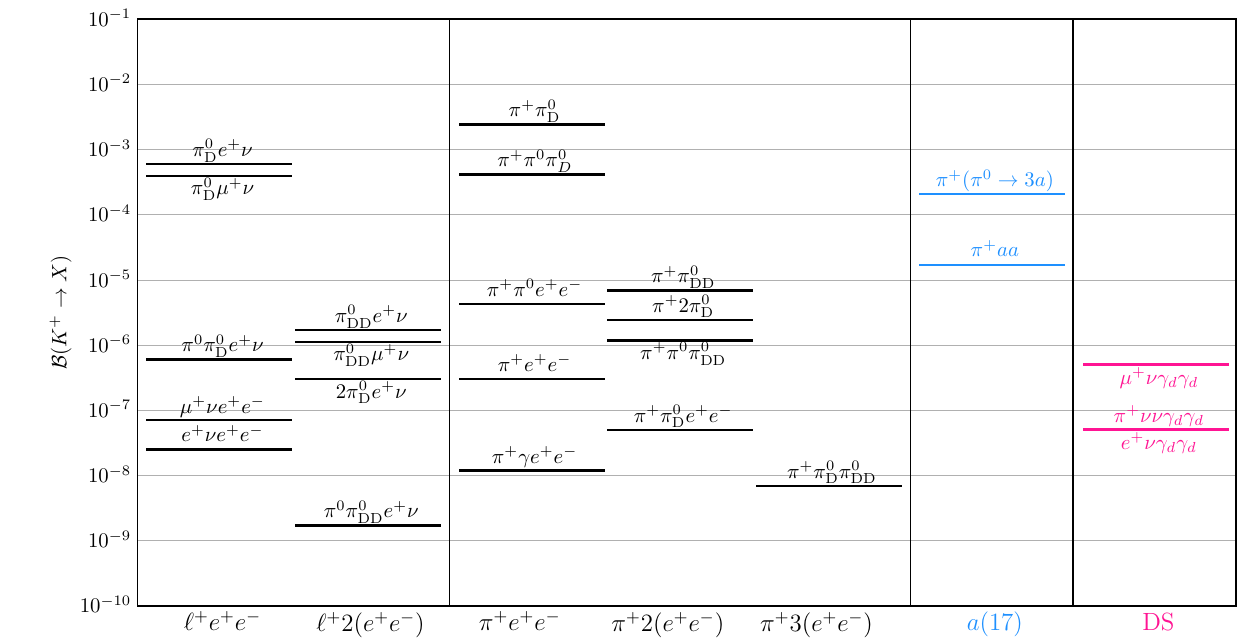}
    \caption{The branching ratio of $K_L$ (top) and $K^+$ (bottom) to multi-lepton final states. In black, we show the SM decay modes~\cite{10.1093/ptep/ptaa104} for each relevant multi-lepton final state. The list is exclusive in the charged particle content but inclusive on the neutral particle content. In blue (second to last column), we show the new physics prediction of a $17$~MeV axion, and in violet (last column) we show the large rates that can be obtained for specific benchmark points in the dark sector models.}
    \label{fig:BRs_bkg}
\end{figure*}

Another natural possibility to consider is the exclusive direct production of two dark bosons. In this case, only neutral mesons are relevant,
\begin{align}\label{eq:4e}
    K_{S,L} & \to X \, X \to 2(e^+e^-).
\end{align}
These decay modes have received some interest due to the possibility to measure sgn$\left[\mathcal{M}(K_L\to\gamma\gamma)\right]$, a stringent test of the Cabibbo–Kobayashi–Maskawa matrix~\cite{Isidori:2003ts,DAmbrosio:2013qmd}.
Only the channel $K_L\to 2(e^+e^-)$ has received experimental attention to date. The most precise measurement was performed by KTEV~\cite{AlaviHarati:2001ab} with a total of 441 events,
\begin{equation}\label{eq:k4e_KTEV}
    \mathcal{B}\left(K_L \to 2(e^+e^-)\right)_{\rm KTEV}
    = (3.72\pm0.29)\times 10^{-8},
\end{equation}
where the total error quoted is the sum in quadrature of statistical and systematical errors. The measurement by NA48~\cite{Lai:2005kw} is compatible but was based on a smaller sample of 132 events, $\mathcal{B}\left(K_L \to 2(e^+e^-)\right)_{\rm NA48}
    = (3.67\pm0.40)\times 10^{-8}$.
Both measurements are in agreement with the SM prediction~\cite{Zhang:1997et} of $\mathcal{B}\left(K_L \to 2(e^+e^-)\right)_{\rm SM} = 3.65\times10^{-8}$ from Ref.~\cite{DAmbrosio:2013qmd}. KTEV provides event distribution plots in $x=(m_{ee}/m_K)^2$ as a double entry histogram with bin size $\Delta x = 0.1$~\footnote{KTEV further processes their data with cut on $m_{ee}>8$~MeV for both pairs when measuring the $CP$ violating amplitude. This reduces the total number of events to 264, but no invariant mass distribution is available in that case.}. 

For the $K_S\to 2(e^+e^-)$ decays no search exists and the SM rate is expected at the level of $\mathcal{O}(10^{-10})$~\cite{DAmbrosio:2013qmd}. Recently, the possibility to measure this decay at LHCb has been discussed~\cite{MarinBenito:2193358,Junior:2018odx}, where the single event sensitivity to the similar decay mode $K_S\to\pi^+\pi^- e^+e^-$ is expected to be below $\mathcal{O}(10^{-9})$~\cite{Dettori:2019oak} per fb$^{-1}$. Clearly, sensitivities at this level for the corresponding four-lepton mode would represent a significant improvement given the absence of previous searches. Other experiments to be considered are KLOE and KLOE-2, which collected data from 2001 to 2006 and from 2014 to 2018, respectively. Over $10^{9}$ $K_S$ states were produced in $\phi\to K_L K_S$ decays~\cite{Ambrosino:2008zi} and the detector was also used to search for multi-lepton final states in the double-Dalitz decays of the $\eta$ meson~\cite{KLOE2:2011aa}. 

Finally, $X$ pairs can also be produced alongside missing energy. Foreshadowing our example with HNLs, it is possible for $X$ to be produced in a cascade of the type $X^\prime \to NN \to \nu X \nu X$, motivating searches for
\begin{align}\label{eq:pi02nu2ee}
    K_{S,L} &\to \pi^0 2(\nu X) \to \pi^0 \, 2(\nu e^+e^-),
    \\\label{eq:piplus2nu2ee}
    K^+ &\to \pi^+ 2(\nu X) \to \pi^+ \, 2(\nu e^+e^-),
\end{align}
which, despite not allowing for a full reconstruction of the kaon mass, can still be used to search for the $X$ resonance. Note that in this case $N$ does not need to mix with neutrinos, and could be an arbitrary dark fermion that decays to a lighter invisible state while emitting $X$. If it does mix with electron- or muon-neutrinos, then $N$ can also be produced in charged-current kaon decays, leading to a striking five-lepton final state via $N\to \nu X^\prime\to\nu XX$. Generically, this signature is simply
\begin{align}\label{eq:ellnu2ee}
    K^+ &\to \ell^+ \nu XX \to \ell^+\nu \, 2(e^+e^-),
\end{align}
where $\ell \in \{e,\mu\}$. Since $|P_K - P_{\pi}|$ can be measured on an event-by-event basis in charged kaon decays (\ref{eq:piplus2nu2ee}) and (\ref{eq:ellnu2ee}), the intermediate dark resonances can be searched for as an excess of events around $|P_K - P_{\pi}| = m_{X^\prime}$ and $|P_K - P_{\ell}|=m_N$, respectively.

Finally, we note that the channels $K_{S,L}\to NN\to2(\nu X) \to 2(\nu e^+e^-)$ are more challenging experimentally and are suppressed in minimal models with only Higgs portal or kinetic mixing, so we do not consider them here. It is, however, possible that these rates are large in models with mass mixing between the SM $Z$ and a new $Z^\prime$ boson, as discussed in Ref.~\cite{Hostert:2020gou}.

\subsection{Backgrounds}

The main SM backgrounds to multi-lepton final states in kaon decays arise from Dalitz decays of $\pi^0$. Therefore, the probability to mimic a multiplicity of $n$ $(e^+e^-)$ pairs decreases with larger $n$, as it relies on double Dalitz, $\pi^0_{\rm DD}\to 2(e^+e^-)$, or multiple single Dalitz decays, $\pi^0_{\rm D}\to \gamma e^+e^-$. We illustrate this effect in \reffig{fig:BRs_bkg}, where all SM decays containing a fixed number of $(e^+e^-)$ pairs are shown. We do not show modes with additional charged particles but include modes with additional neutral particles. 

The power to reject the backgrounds in \reffig{fig:BRs_bkg} will depend crucially on the experimental efficiency to detect and veto additional neutral particles produced alongside the desired final state. This is most important for $\pi^0$ Dalitz decays, where the photon can be emitted at large angles and be very soft. The Dalitz rate, however, can also be reduced by requiring that $m_{ee}$ is either larger or \emph{smaller} than $m_{\pi}$. For instance, NA62 was able to reach sensitivities of $\mathcal{O}(10^{-10})$ in the lepton-number violating $K^+\to\pi^- e^+e^+$ decays~\cite{CortinaGil:2019dnd}, and, in the same analysis, conclude that the SM decays $K^+\to\pi^+e^+e^-$ were observed with a negligible background in the region of $m_{ee}<100$~MeV. 

Additional charged pions are more effectively vetoed than photons unless they decay in flight, which was found to be a negligible contribution in~\cite{CortinaGil:2019dnd}. Even when accounting for $\pi^\pm\leftrightarrow e^\pm$ mis-identification, such decay modes can be rejected by correctly reconstructing the kaon invariant mass in visible channels, or by correctly reconstructing the missing mass in channels with a single invisible particle in the final state. The same technique would also reduce backgrounds from $\mu^\pm \leftrightarrow e^\pm$ mis-identification.

\section{Dark cascades in a $U(1)_d$ dark sector}
\begin{figure*}[t]
    \centering
    \includegraphics[width=\textwidth]{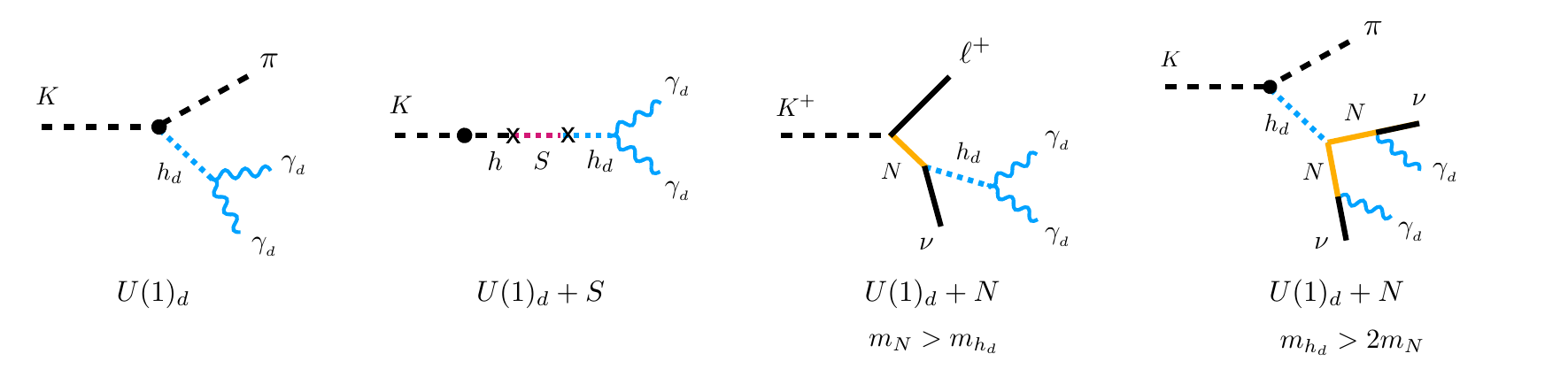}
    \caption{Kaon decay diagrams in minimal and extended $U(1)_d$ dark sectors that lead to observable $2(e^+e^-)$ signatures.}
    \label{fig:DS_diagrams}
\end{figure*}

We will start by studying a secluded $U(1)_d$ gauge symmetry extension of the SM, broken at the MeV scale. More specifically, we work with three different scenarios, starting with a minimal $U(1)_d$ model containing only a dark photon and a dark scalar, and later we extend it by either an additional scalar singlet $S$ or by new heavy neutrino fields. In kaon decays, we will be interested in Higgs and neutrino portal production of dark states and will neglect production through kinetic mixing, which is not particularly advantageous in the case of kaon decays~\cite{Pospelov:2008zw}.
Nevertheless, kinetic mixing will still be large for $X\to e^+e^-$ decays to be prompt. The decay channels of interest are shown in \reffig{fig:DS_diagrams}.

The minimal $U(1)_d$ DS is described by the Lagrangian
\begin{align}\label{eq:higgsedU1}
    \mathcal{L}_{\rm DS} &= |D^\mu \phi|^2 - \frac{1}{4} F_{d}^{\mu \nu}F_{d\,\mu \nu}  - \frac{\varepsilon}{2} F_{d}^{\mu \nu} F_{\mu\nu} \\ 
    & \quad -\mu^2(\phi^\dagger \phi) - \lambda(\phi^\dagger \phi)^2 - \lambda_d (\phi^\dagger \phi)(H^\dagger H), \nonumber
\end{align}
where $D^\mu\equiv \partial^\mu - i g_d \widehat{\Adark}^\mu$ and $F_{(d)}^{\mu\nu}$ is the (dark) photon field strength tensor. The scalar $\phi$ spontaneously breaks the $U(1)_d$ symmetry with a vacuum expectation value $\langle \phi \rangle \equiv v_d$. After the symmetry is broken, we write $\phi = (\widehat{\hdark} + v_d)/\sqrt{2}$ and are left with the physical dark scalar $\hdark$ and the physical dark photon $\Adark$. In the limit of small portal couplings, their masses are fully specified by $m_{\hdark} = \sqrt{2\lambda} v_d$ and $m_{\Adark} = g_d v_d$.

We will focus on the case where the mass spectrum satisfies $m_{\hdark}>2m_{\Adark}$, so that once the dark scalar is produced, it can undergo prompt $\hdark \to \Adark\Adark$ decays. The dark photons then decay electromagnetically with typical lifetimes of
\begin{equation}
    c\tau^0_\Adark \simeq 70\,\mu\text{m}\left( \frac{2\times10^{-4}}{\epsilon} \right)^2 \left( \frac{30 \text{ MeV}}{m_\Adark} \right).
\end{equation}
In the minimal dark photon model (without a dark higgs), this regime of $\mathcal{O}(10$s$-100$s$)$~MeV masses and $\mathcal{O}(10^{-5}-10^{-4})$ kinetic mixing is challenging to probe experimentally, as the dark photon is very weakly coupled but still short-lived. This naturally avoids existing limits set by beam dump, fixed target, and collider experiments (see Ref.~\cite{Fabbrichesi:2020wbt} for a recent review). In addition, dark photons produced in supernovae do not escape the core and have their energy reabsorbed, avoiding a prohibitively large cooling of the supernova~\cite{Chang:2018rso}. While experimental progress in this region of parameter space can be expected from future searches at LHCb~\cite{Ilten:2015hya,Ilten:2016tkc},  Belle-II~\cite{Belle-II:2018jsg}, HPS~\cite{HPS:2018xkw,Bravo:2019yuw} and APEX~\cite{APEX:2011dww} at J-Lab, MESA at Mainz~\cite{Doria:2019sux}, and Mu3e~\cite{Echenard:2014lma} at PSI, one can already derive constraints on such short-lived dark photons in more involved models. For instance, in the presence of a dark Higgs or new scalar portal particles, we find that several meson decays become competitive probes of short-lived dark photons.

\subsection{Minimal model}\label{sec:KtopiX_minimal}

\subsubsection{$K\to \pi 2(e^+e^-) $}
Decays of the type $K\to \pi \hdark$ are well known probes of light scalars mixed with the Higgs. The rate was calculated for a light Higgs boson~\cite{Leutwyler:1989xj}, from which one can derive the result for a light scalar mixed with the Higgs~\cite{Bezrukov:2009yw},
\begin{align}\label{eq:KtopiS}
    \Gamma_{K_L\to\pi^0 \hdark} &= \frac{ (s_\theta \Re(g_{sd}))^2}{64\pi} \frac{m_K^3m_h^4}{v^2} \lambda^{1/2}(1,r_\pi^2, r_\hdark^2),
    \\
    \Gamma_{K^+\to\pi^+ \hdark} &= \frac{ s_\theta^2 |g_{sd}|^2}{64\pi} \frac{m_K^3m_h^4}{v^2} \lambda^{1/2}(1,r_\pi^2, r_\hdark^2),
\end{align}
where $g_{sd}$ is the SM FCNC coefficient in $s\to d$ transitions,
\begin{equation}\label{eq:gsd}
g_{sd} = \frac{3}{32\pi^2m_h^2}(y_t^2V_{ts}^*V_{td}+y_c^2V_{cs}^*V_{cd}+y_u^2V_{us}^*V_{ud}),
\end{equation}
dominated by the top quark contribution. For values of allowed $s_\theta$ scalar mixing, the subsequent cascade in the dark sector allows us to predict
\begin{align}\label{eq:BRKtopiX}
    \mathcal{B}\left({K_L\to\pi^0 2(e^+e^-)}\right) &\simeq 2 \times 10^{-7} \left(\frac{s_{\theta}}{5\times 10^{-3}}\right)^2,
    \\\label{eq:BRKtopiX2}
    \mathcal{B}\left({K^+\to\pi^+ 2(e^+e^-)}\right) &\simeq 5 \times 10^{-8} \left(\frac{s_{\theta}}{5\times 10^{-3}}\right)^2.
\end{align}

Note that beam dump and $K\to\pi X$ searches for the dark scalar do not apply in this case, as its decays are prompt and into 4 leptons. Still, the mixed-quartic term is constrained by decays of the Higgs. Requiring that $\mathcal{B} (h \to\hdark \hdark) < 10\%~$, amounts to $\lambda_{\Phi H} \lesssim 10^{-2}$, which in terms of the mixing angle is simply $\theta \lesssim 8\times10^{-5}\times(v_d/500~\text{MeV})$. Satisfying this bound and accommodating values of $\theta$ as large as the one used in \refeqs{eq:BRKtopiX}{eq:BRKtopiX2} requires a large vev of $v_d \sim 30$~GeV, which would still predict that $\hdark$ is short-lived ($c \tau^0_{\hdark} \lesssim 10^{-5}$~cm for $m_{\hdark} = 100$~MeV and $m_{\Adark}=30$~MeV). 

It is also possible to search for the visible decays of pairs of dark photons or scalars produced in Higgs decays. These multiple lepton final states lead to lepton jets at the LHC~\cite{Gopalakrishna:2008dv,Falkowski:2010cm,Curtin:2014cca,Curtin:2013fra,Davoudiasl:2013aya}, which have been searched at ATLAS~\cite{ATLAS:2014fzk,ATLAS:2015itk,ATLAS:2021ldb} and CMS~\cite{CMS:2021pcy} at larger dark photon masses.  We do not find any limits in the region $m_{\Adark} < 100$~MeV, but we note that the current model predicts branching ratios for two-lepton as well as four-lepton jet final states as large as $10\%$. At large masses, these searches place constraints on $\lambda_{\Phi H}$ as strong as $\lambda_{\Phi H}<10^{-4}$, which, if extended to lower masses, are complementary to the kaon decay searches, which are sensitive to $\theta$ instead.

\subsection{Singlet-scalar extension}

As a direct application of our previous work~\cite{Hostert:2020gou}, one can compute the BR for $K\to(\hdark)^*\to\Adark\Adark$ in the minimal model of \refeq{eq:higgsedU1} to find that current constraints on the SM Higgs decays limit $\lambda_d$ to a small value and require $\mathcal{B}(K_L\to\Adark\Adark) \lesssim 10^{-17}$. This suppression is a consequence of the hierarchy of scales appearing in $K$ decays, ($m_K$/EW)$^4$, when compared to $h\to\gamma_d\gamma_d$ decays (EW/EW)$^4$.
Here EW stands for a heavy mass scale, such as $m_h$, $m_W$, etc. This scaling, however, is not a generic feature of Higgs portal models. For instance, if in addition to the dark Higgs we also allowed for a real scalar $S$ with super-renormalizable couplings to the SM and DS Higgses, then kaon and Higgs decays to the DS would both proceed via exchange of virtual $S$, and kaon decays would be suppressed by $m_K^4/(m_S^2 \times $EW$^2$) instead. 

More specifically, we consider a trivial extension of \refeq{eq:higgsedU1} by
\begin{equation}
    \mathcal{L}_{S} = \partial_\mu S\partial^\mu S - m_S^2 S^2 -A_H H^\dagger H S - A_{\phi} \phi^\dagger \phi S.
\end{equation}
Under the hierarchy of $v_d\ll m_S \ll v$, the small mixing angles between the scalars are simply $\theta_{S h} \simeq  A_H /(2\lambda v)$ and $\theta_{S \hdark} \simeq A_\phi v_d / (m_{\hdark}^2 - m_S^2)$. 

We are particularly interested in singlet scalars above the kaon mass, $m_S \gtrsim 500$~MeV, which do not violate constraints from Higgs decays. Requiring that $\mathcal{B}(H\to \gamma_d \gamma_d) < 10\%$ for $m_S=1$~GeV, $m_{\hdark}=100$~MeV, we find $A_\phi A_H \sim 10$~MeV$^2$. Note that such values are natural, in the sense that corrections to $m_S$, $m_\phi$, and $A_{\phi, H}$ themselves, are never too large in the region of interest~\cite{DiLuzio:2016sur}.

\subsubsection{$K\to2(e^+e^-)$}
We compute the rate for $K_L$ as well as $h$ decays to dark photons, leaving the details to \refapp{app:calculation_rate}, where for the $CP$-odd $K_L$ state, we find
\begin{align}\label{eq:KLto2Adark}
    \mathcal{B}({K_L\to \Adark \Adark}) &\simeq 5\times10^{-8} \times  \left(\frac{\mathcal{B}({h\to \Adark \Adark})}{10\%}\right)
    \\\nonumber
    &\times\left( \frac{m_K^4}{(m_K^2-m_S^2)^2 + m_S^2\Gamma_S^2}\right),
\end{align}
which for a narrow resonance with $m_{S} = 1$~GeV, takes values of $5\times10^{-9}$. For the decays of $CP$-even $K_S$ states, we find an enhancement of the rate by $\approx(\Re{V_{ts}^*V_{td}}/\Im{V_{ts}^*V_{td}})^2\approx 5.2$, which corresponds to
\begin{align}
    \mathcal{B}(K_S\to\gamma_d\gamma_d) &\simeq 5.2\times \left(\frac{\tau_{K_S}}{\tau_{K_L}}\right) \times \mathcal{B}(K_L\to\gamma_d\gamma_d)
    \\
    &\simeq 9\times 10^{-3}\times \mathcal{B}(K_L\to\gamma_d\gamma_d). \nonumber
\end{align}
The rate in \refeq{eq:KLto2Adark} is enhanced when $m_{S} \sim m_{K}$, but should not exceed the measurement in \refeq{eq:k4e_KTEV}. A dedicated search for visible resonances would improve constraints in the current scenario.

We note also that decays to muons can also contribute in this case as $m_K > 4 m_\mu$. In that case, the branching ratio for $K\to (\Adark \to e^+e^-)(\Adark \to \mu^+\mu^-)$ will be larger than the four muon case and will be severely constrained by the KTEV measurement,  $\mathcal{B}(K_L \to e^+e^-\mu^+\mu^-)|_{\rm KTEV} = (2.69 \pm 0.27)\times 10^{-9}$~\cite{KTeV:2002kut}, where we summed statistical and systematic errors in quadrature.

\subsection{Missing energy from a dark neutrino sector}

All cases considered thus far have involved only bosonic dark particles produced via Higgs portal couplings. An additional possibility is the production of the DS through the neutrino portal, where missing energy will be necessarily present. A minimal model is
\begin{align}\label{eq:nudL}
    \mathcal{L}_{\rm DS-\nu} &= \overline{N_R}i\slashed{\partial}N_R + \overline{N_L}i\slashed{D}N_L 
    \\\nonumber
    &\quad + y_d \overline{N_L}\phi N_R + y_\alpha \overline{L}_\alpha \widetilde{H} N_R,
\end{align}
where $N_R$ is a complete singlet fermion, and $N_L$ is a dark fermion charged under the $U(1)_d$. The $U(1)_d$ gauge anomalies may be canceled by an additional and oppositely charged fermion $N^\prime_L$, which together with its singlet partner $N_R^\prime$, may be decoupled from the SM in the limit of vanishing $y_{\alpha}^\prime$ portal coupling. In addition, we will not be concerned with light neutrino masses and assume lepton number is conserved. Relaxing the previous assumptions amounts to considering the Majorana mass terms for the singlet states and $\overline{N^c_L} N_L^\prime$, potentially enriching the dark sector spectrum and vertices but leaving our main message unaltered. Under the previous simplifying assumptions, the neutrino mass matrix can be diagonalized in the broken phase to find massless and mostly-active neutrinos $\nu_{i}$, $i=1,2,3$, and a new Dirac HNL $\nudark$, with a mass $m_\nudark=y_d v_d/\sqrt{2}$. The mixing of $\nudark$ with active neutrinos is simply $|U_{\alpha 4}|^2 = y_\alpha v / y_d v_d$, which we treat as a free parameter.

\subsubsection{$K^+\to \ell_\alpha^+ \nu 4e$}

This kaon decay channel corresponds to a five-lepton final state and the decay chain is kick started by a charged-current process involving the neutrino portal. It is well-known that the rate for $K^+\to \ell_\alpha^+ \nudark$ is enhanced in the presence of massive neutrinos due to the lifting of helicity suppression, and will be controlled by the neutrino mixing parameter~\cite{Shrock:1980vy,Shrock:1980ct},
\begin{equation}
    \mathcal{B}(K^+\to\ell_{\alpha}^+ N) \simeq   |U_{\alpha 4}|^2 \mathcal{B}(K^+\to\ell_{\alpha}^+\nu) \rho(r_\ell,r_N),
\end{equation}
where $\rho(x,y)=\lambda^{1/2}(1,x,y)\left((x+y)-(x-y)^2\right)/(x(1-x)^2)$ is a kinematic function. The constraints on the mixing angle vary between the $\alpha=e$ and $\alpha=\mu$ cases, where the latest HNL searches by NA62~\cite{CortinaGil:2017mqf,NA62:2020mcv} in $K\to\ell_{\alpha} N$ constrain $|U_{e4}|^2\lesssim10^{-9}$ and $|U_{\mu4}|^2 \lesssim 10^{-8}$. In reality, these bounds are somewhat relaxed by factors of at least $\mathcal{O}(100)$ due to the visible decays of $\nudark$, so we identify the values of $|U_{e(\mu)}|^2 = 10^{-7(-6)}$ as a conservative estimate for the upper bound on the mixing angles for $m_\nudark < 350$~MeV. In addition, a rough estimate of neutrino-electron scattering constraints~\cite{Arguelles:2018mtc} can be found to be $|U_{\mu 4}|^2 \epsilon^2 \lesssim 4 \times 10^{-10}$ for dark photons light dark photons, $m_\Adark \ll 100$~MeV and $m_{\nudark} \ll 300$ MeV.

Once $N$ is produced, the dark cascade develops as $\nudark \to \nu \Adark$ if $m_\Adark < m_\nudark$, leading to $K^+\to\ell_\alpha^+\nu e^+e^-$ decays discussed in \refref{Ballett:2019pyw}. However, if the dark scalar is also a kinematically viable final state, e.g. $2m_{\Adark} < m_{\hdark} < m_\nudark$, then the HNL decays to dark photon compete with those into dark Higgs. The corresponding rates are,
\begin{align}
    \Gamma_{\nudark \to \nu \Adark} &= |U_{D 4}|^2 \sum_{i=1}^{3} |U_{D i}|^2 \frac{\alpha_D}{4} \frac{m_4^3}{m_{A^\prime}^2}
    \left(1-r\right)^2 \left(1 + 2 r\right),
    \\ 
    \Gamma_{\nudark \to \nu \hdark} &= |U_{D 4}|^2 \sum_{i=1}^{3} |U_{D i}|^2 \frac{y_d^{2}}{32\pi} {m_\nudark}\left(1-r\right)^2,
\end{align}
where $r=m_\Adark^2/m_\nudark^2$. In our minimal model the typical branching ratios for these two decays are comparable by virtue of $2 g_d^2 m_N^2/m_{\Adark}^2 = y_d^2$. Neglecting the final state masses, it is then safe to assume that $\mathcal{B}(N\to \nu \hdark) \simeq 50 \%$. Note that it is also possible for $\nudark\to\nu\Adark$ to be forbidden or strongly suppressed with respect to decays to $\hdark$ in more complete models. For instance, in Ref.~\cite{Abdullahi:2020nyr} a left-right symmetry in the dark neutrino sector forbid certain $\nudark-\nu-\Adark$ vertices, but not $\nudark-\nu-\hdark$. Regardless of the model, if the dark cascade proceeds as follows
\begin{equation}
    K^+\to \ell_\alpha \left(N\to \nu \hdark \to \nu \gamma_d \gamma_d \to \nu 2(e^+e^-) \right),
\end{equation}
then \emph{all} dark sector resonances of our minimal model can be found in this decay channel by searching for an excess of events with the following four relations on the kinematics,
\begin{align}
    m_{ee} &\mbeq m_\Adark, 
    \\
    m_{ee}^\prime&\mbeq m_\Adark,
    \\
    m_{4e}&\mbeq m_\hdark,
    \\
    m_{\nu4e}=|p_K-p_{\ell_\alpha}| &\mbeq m_\nudark,
\end{align}
in addition to a vanishing missing mass $m_\nu = |p_K-p_\ell-p_{4e}| \to 0$. Moreover, due to the 2-body nature of the decays, kinematics alone constrains two other relevant invariants, $m_{\alpha \nu}^2 = (p_K - p_{4e})^2$ and $m_{\alpha 4e}^2 = (p_\alpha + p_{4e})^2$, so that for every event the relation
\begin{equation}
    m_{\alpha \nu}^2+m_{\alpha 4e}^2+m_{\nu4e}^2 = m_{K}^2 + m_{\alpha}^2 + m_{\nu}^2 + m_{4e}^2
\end{equation}
is satisfied. The kinematical range of $m_{\alpha \nu}^2$ and $m_{\alpha 4e}^2$ may be adapted from~\cite{Ballett:2019pyw}, and cuts on $m_{\alpha \nu}>m_{\pi}$ can help further reduce backgrounds from misidentification or fast pion decay, e.g. in $K^+\to (\pi^+\to \ell^+\nu)2(e^+e^-)$.

\subsubsection{$K \to \pi 2(\nu e^+e^-)$}

As an additional possibility, we want to point out a more challenging decay chain that appears when $m_\Adark < m_N < m_\hdark/2$, namely
\begin{align}
    K_{S,L}&\to\pi^0 \left(\hdark \to \nudark\nudark \to 2(\nu \Adark) \to 2(\nu e^+e^-)\right),
    \\
    K^+&\to\pi^+ \left(\hdark \to \nudark\nudark \to 2(\nu \Adark) \to 2(\nu e^+e^-)\right).
\end{align}
Including production and the requirement of fast decays of the final state dark states, this decay mode requires that every one of the three portals couplings be sizable. Even if all decays are prompt, the rate will be typically smaller than the estimate in \refeq{eq:BRKtopiX} as $\mathcal{B}(h_d\to \nudark\nudark) < \mathcal{B}(\hdark \to \Adark \Adark)$. In particular, in our minimal model, the rate for these two processes are given by
\begin{align}\label{eq:hdarkdecays}
    \Gamma_{\hdark\to\Adark\Adark} &= \frac{\alpha_D}{16}\frac{m_\hdark^3}{m_{\Adark}^2} f(r_\hdark),
    \\
    \Gamma_{\hdark\to\nudark\nudark} &= \frac{y_d^2 m_\hdark}{16 \pi} (1-4r_\hdark^2)^{3/2},
\end{align}
where $f(r_P)=(1 - 4 r_P^2 + 12 r_P^4)\lambda^{1/2}(1,r_P^2,r_P^2)$ with $r_P=m_{\Adark}/m_P$. Neglecting final state masses and any additional decay channel, one may estimate the typical BR as 
\begin{equation}
    \mathcal{B}(\hdark \to \nudark\nudark) \simeq \frac{1}{1 + \frac{m_\hdark^2}{8m_\nudark^2}} \simeq 10\% \times \left(\frac{8 m_\nudark}{m_\hdark}\right)^2.
\end{equation}
This is pushing the overall branching ratio for $K\to\pi2(\nu e^+e^-)$ to values of $\mathcal{O}(10^{-9})$ and below. Experimentally, the situation is much alike that already explored in \refsec{sec:KtopiX_minimal}, with the added challenge that the parent kaon mass is not reconstructed on an event-by-event basis. For charged kaons, the dark Higgs resonance can be found as a bump in $m_{2\nu4e} = |p_K - p_\pi| \mbeq m_\hdark$, although, in this case, the $N$ mass is not fully reconstructed. The latter may be inferred statistically by searching for a cut-off in the lepton pair invariant masses, since $m_{ee},m_{ee}^\prime < m_N$. In conclusion, $K\to\pi2(\nu e^+e^-)$ is likely not competitive with $K\to \pi 2(e^+e^-)$, but it would offer additional discriminatory power in the case of a signal in the fully visible channels.

\subsection{Higher $(e^+e^-)$ multiplicity}

The previous discussion showed how the multiplicity of lepton pairs in kaon decays can rapidly grow due to fast decay chains in the dark sector. We now take this argument to a more extreme extent and point out more exotic possibilities. For example, in addition to the double production of dark photons considered above, the singlet scalar and HNL extensions of the dark $U(1)_d$ model allow also for triple and quadruple production of $\Adark$. These channels often compete with lower $(e^+e^-)$ multiplicity modes but can provide alternative probes of the DS models we discuss.

Already in the $U(1)_d+S$ model we consider, the fragmentation of the singlet scalar $S$ to 
two dark Higgses to four dark photons may result in sizable rates for 
\begin{equation}
    K^+\to \pi^+ \left(S \to \hdark \hdark\to 
\Adark \Adark \Adark \Adark \to 4(e^+e^-)\right),
\end{equation}
giving a signature of $K^+$ decay with nine charged tracks. Similar cascade decay will then exists for the $K_L$, $K_L\to \pi^0 4(e^+e^-)$. The rates are still given by \refeq{eq:KtopiS}, where now $\theta \to \theta_{Sh}$. One advantage of such decays is that $\theta_{Sh}$ is not constrained by SM Higgs decays, and may be much larger than $\theta$. In the absence of dedicated experimental searches, it is difficult to speculate how large a branching these decays may have within existing datasets.

The dark neutrino sector in \refeq{eq:nudL} also admits, and to a certain extent calls for, the existence of at least a pair of HNLs. If both of these states are kinematically accessible below the kaon mass and heavier than the dark bosons, then several inter-generational decay chains can take place. With two generations, one already expects decays with 5, 7, and 9 charged leptons,
\begin{align}
    K^+&\to\ell^+ \left( N^\prime \to N \Adark \to \nu \Adark\Adark \to \nu 2(e^+e^-) \right),
    \\
    K^+&\to\ell^+ \left( N^\prime \to N \Adark \to \nu \Adark\hdark \to \nu 3(e^+e^-) \right),
    \\
    K^+&\to\ell^+ \left( N^\prime \to N \hdark \to \nu \hdark\hdark \to \nu 4(e^+e^-) \right),
\end{align}
where all the dark resonance masses are, in principle, measurable. Similar decay chains can proceed via scalar portal decays $K\to\pi \hdark \to \pi N^\prime N^{(\prime)}$ pairs, with yielding up to four $(e^+e^-)$ pairs. The latter decays are expected in more generic DS models with multiple generations of fermions that may or may not mix with neutrinos, and which are heavier than the dark bosons. 

Other multi-lepton final states from dark higgstrahlung, $\Adark^* \to \Adark \hdark$, are also unavoidable in such DS models, but do not appear at interesting branching ratios (see \refapp{app:higgstrahlung}).

\section{Multiple production of MeV-scale axions}

\begin{figure}[t]
    \centering
    \includegraphics[width=0.44\textwidth]{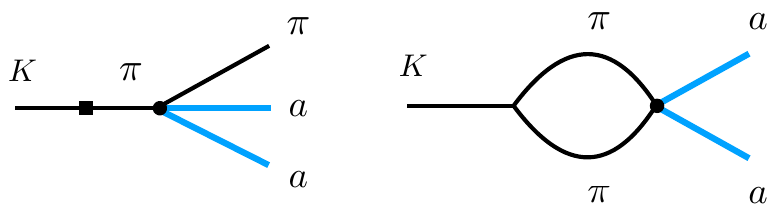}
    \caption{Diagrams of kaon decay probes of the 17 MeV axion. We show the two multi-lepton kaon decay signatures arising from the $a-a-\pi^0-\pi^0$ vertex.\label{fig:axion_diagrams}}
\end{figure}
\begin{figure}[t]
    \centering
    \includegraphics[width=0.42\textwidth]{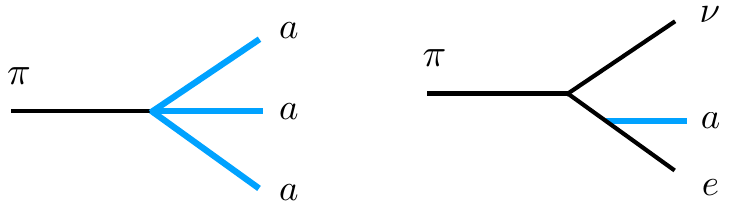}
    \includegraphics[width=0.3\textwidth]{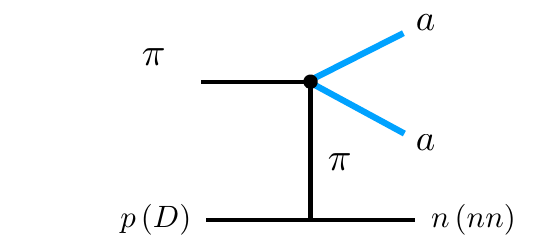}
    \caption{Diagrams of pion probes of the 17 MeV axion. On the top row, we show the two rare pion decay signatures, the left one arising from the $a-a-a-\pi^0$ vertex and the right one from the coupling of the $a(17)$ to electrons. On the bottom, we also show the double axion production in pion capture by free protons (p) or deuterons (D). \label{fig:axion_diagrams2}}
\end{figure}

The QCD axion provides an elegant solution to the strong $CP$ problem in the SM. In general terms, the axion is the pseudo-Nambu-Goldstone (pNG) boson of a global $U(1)_{PQ}$ Peccei-Quinn (PQ) symmetry, spontaneously broken at a scale $f_a$.
In minimal models, the QCD axion mass is generated via non-perturbative effects as $m_a \propto m_\pi F_\pi/f_a$. 

Even though most variations of the PQ mechanism with $m_a$ in the MeV scale are excluded, Refs.~\cite{Alves:2017avw,Alves:2020xhf} propose that axions that are \emph{pio-phobic} and \emph{muonphobic} are still allowed. This is mostly due to the uncertainties in the axio-hadronic couplings as obtained from chiral perturbation theory (ChPT). While building a self-consistent and UV-complete model for such an axion that would pass all experimental constraints and theoretical consistency checks remain a challenge, the model is worth addressing, as it may give an explanation to some intriguing experimental anomalies.

In particular, Ref.~\cite{Alves:2020xhf} speculates if MeV-scale axions are \emph{electrophilic} and have a mass of $16.8$~MeV, they are able to explain the anomaly in the $e^+e^-$ angular spectrum observed in $^8$Be nuclear de-excitation by the ATOMKI experiment~\cite{Krasznahorkay:2015iga}, at a significance of $6.8\sigma$. This explanation, according to Ref.~\cite{Alves:2020xhf}, is also corroborated by similar observations in $^4$He nuclear transitions at a significance of $7.2\sigma$~\cite{Krasznahorkay:2019lyl} and by a longstanding $2-3.2\sigma$ discrepancy in the KTEV $\pi^0\to e^+e^-$ measurement~\cite{Abouzaid:2006kk}. While the explanation for such experimental anomalies may very well be mundane, their $a(17)$ interpretation will serve to us as a motivation to consider the impact of this particle on kaon decays. As we will show, if such particle exists, it would lead to extraordinary signatures in $K\to (\pi) 4e$ and $K\to 2\pi \,6e$ data, with a double and triple coincidence in $e^+e^-$ invariant masses, respectively.

The couplings of $a(17)$ to matter are determined by the PQ charges of the SM fermions. In the IR, one can proceed to compute observables by considering the axion as an axial rotation on the fermion masses, $m_\psi \to m_\psi e^{i \gamma_5 q_{PQ}^\psi a/f_a}$, and following the assumption of \emph{pio-phobia}, we set $q_{PQ}^d/2 = q_{PQ}^u = 1$ and $m_u = m_d/2$, and keep $q_{PQ}^e$ floating. The axion decays predominantly via $a\to e^+e^-$ with a lifetime of 
\begin{equation}
c \tau^0_{a} \simeq \frac{1.2 \, \mu\text{m} }{(q^e_{PQ})^2} \left(\frac{17\text{ MeV}}{m_a}\right) \left(\frac{f_a}{1.03\text{ GeV}}\right)^2
\end{equation}
which can be safely assumed to be prompt for all kaon experiments. 

In addition to the axion coupling to electron, we point out an axio-hadronic vertex that is an unavoidable consequence of the generation of the axion mass from the vacuum condensate. Expanding the chiral Lagrangian in the axion field, we find the vertices
\begin{align}\label{eq:coupling_pi2a2}
    \mathcal{L}_{aa\pi\pi}\supset &\frac{m_{a}^2}{4F_\pi^2} aa \pi^0\pi^0 + \frac{m_{a}^2}{2F_\pi^2} aa \pi^+\pi^-
    +\mathcal{O}\left(\frac{m_{q}^2}{f_a^2},\frac{F_\pi^2}{f_a^2}\right),
\end{align}
where $m_q$ are light quark masses and $F_\pi \simeq 93$~MeV. Therefore, the $a-a-\pi-\pi$ vertex is simply given by 
\begin{equation}\label{eq:aapipivertex}
\frac{m_a^2}{F_\pi^2} = 0.033\times \left(\frac{m_a}{17\,{\rm MeV}} \right)^2 ,
\end{equation}
which is a relatively mild suppression in the case of $a(17)$ (e.g. if compared to the pion-axion mixing that Ref.~\cite{Alves:2020xhf} takes below 
$10^{-3}$ level). This vertex will give rise to several channels with double axion production, such as those shown in \reffigs{fig:axion_diagrams}{fig:axion_diagrams2}.

\subsection{Double axion production in kaon decays}

\subsubsection{$K\to \pi aa$}

The rate for $K_L\to\pi^0aa$ can be calculated under the assumption of pion pole dominance. This is analogous to the procedure to calculate $K_L\to\gamma\gamma$ decays in the SM, where good agreement with data is observed~\cite{Ma:1981eg,Donoghue:1986ti} (see also the reviews in ~\cite{DAmbrosio:1994fgc,Cirigliano:2011ny}). Assuming leading order ChPT operator responsible for $K\to2\pi$ decays, one extracts the value for the mixing matrix element  $\mathcal{M}_{K_L-\pi^0}=-0.07$~MeV$^2$~\cite{Ma:1981eg,Donoghue:1986ti}, which can be applied to compute the full amplitude,
\begin{equation}
    \mathcal{M}_{K_L\to\pi^0aa} = \frac{\mathcal{M}_{K_L-\pi^0}}{m_{K_L}^2 - m_{\pi^0}^2} \times \frac{m_a^2}{F_\pi^2}.
\end{equation}
In principle, $K-K-a-a$ vertex followed by $K-\pi$ mixing will also contribute, but in the limit $m_K \gg m_\pi$ the emission from the pion line is expected to be dominant. 
A trivial integration over phase space yields,
\begin{align}
    \mathcal{B}(K_L\to \pi^0 a a) \simeq 7\times 10^{-5},
\end{align}
where $m_a = 17$\,MeV was used. 
This is significantly above expectation in the SM and well within reach of experimental searches. In fact, given such large rates, it is not inconceivable that an inclusive search for 4 charged tracks with 6 energy clusters in the electromagnetic calorimeter of KOTO would be sensitive to such values. Additional information on the 4 charged tracks momenta to reconstruct the $a(17)$ masses and reduce $K_L\to\pi^0\pi^0_{\rm D}\pi^0_{\rm D}$ backgrounds (at an expected rate of $\mathcal{B}\sim \mathcal{O}(6\times10^{-5})$) would be important, but not necessarily required, depending on the photon detection inefficiencies.

Generalizing the same calculation to the $K^+\to \pi^+ aa$ decay, and for 
$m_a = 17$\,MeV we find 
\begin{equation}
    \mathcal{B}(K^+\to \pi^+ a a) \simeq 1.7\times 10^{-5},
\end{equation}
where the smaller branching compared to $K_L$ case is simply due to a shorter lifetime of $K^+$. A background to this process may come from {\em e.g.} $K^+\to \pi^+\pi^0_{\rm DD}$ where $\pi^0_{\rm DD}$ stands for $\pi^0$ undergoing double-Dalitz decay to four leptons. The rate for such a process is $\sim 7\times 10^{-6}$, smaller than the expected signals from $2a(17)$ decay mode. It can be further removed with $m_{4l} \geq 150$\,MeV cut. 

The calculations above assume the dominance of $K-\pi$ mixing, which follows from the assumption that leading order ChPT operator $\sim G_8$ dominates the diagrams \cite{Cirigliano:2011ny}. One can question whether this assumption is valid, and how these predictions may change if it is not correct. In connection with $a(17)$, this was recently discussed in Ref.~\cite{Alves:2020xhf}. If for some reason $K-\pi$ mixing diagram is suppressed, there is a way of estimating lower bound on $\mathcal{B}(K\to \pi a a)$ using theoretical analysis developed for {\em e.g.} $K_L\to \pi^0\gamma\gamma$ \cite{Cappiello:1992kk,Cohen:1993ta}. Starting from physical $K\to 3\pi$ amplitudes, one can construct the pion loop induced $K\to \pi aa$. The loop will have the absorptive part, giving a model-independent lower limit on the branching ratio. While analyzing ``small $G_8$" limit is beyond the scope of this work, we note that such absorptive limit will be in the range of ${\cal B}(K_L\to \pi^0aa) \sim 10^{-7}$, which is firmly within current experimental capabilities.

\subsubsection{$K\to aa$}

Yet another smoking gun signature of $a(17)$ are decays of the type $K_{S,L}\to aa$. Again on the theme of large $\pi^2a^2$ interactions, we argue that this decay proceeds primarily through the one pion loop in $K\to\pi\pi\to a a$. Owing to the on-shell $K\to \pi\pi$ decays, the amplitude has both absorptive and dispersive parts. While the calculation of the dispersive part could be somewhat involved due to possible interplay of long- and short distance contributions, intermediate $\sigma$ etc mesons in the long-distance amplitude, the calculation of the absorptive part is in some sense more robust. Also, based on the experience with other decays that develop imaginary part to the amplitude, such as $\pi^0\to e^+e^-$, $K_L \to \mu^+\mu^-$, $\eta\to \mu^+\mu^-$, it is often the case that contribution of the imaginary part to the decay rate is comparable or larger than that of the real part. In the past, the pion loops, giving both real and imaginary parts to the amplitude, were used to derive $K_S\to \gamma\gamma$ decay rates, with satisfactory results, see {\em e.g.} Ref.~\cite{Kambor:1993tv}. 

The absorptive part of the amplitudes factorizes into a loop function times $\mathcal{M}_{K\to\pi^0\pi^0}$, which can be extracted from data, and $\mathcal{M}_{\pi^0\pi^0 \to a a}$, which is calculable, with a twice larger result for a charged pion intermediate state. 
Ignoring possible modifications from the strong re-scattering phases for the 2-pion system,  we get the following estimate for the absorptive amplitude:
\begin{equation}
\mathcal{M}_{K\to aa,{\rm abs}} = \mathcal{M}_{K\to\pi^0\pi^0} \frac{m_a^2}{F_\pi^2} \frac{N_{\pi}}{2} \times I(m_\pi,m_K),
\end{equation}
where $F_\pi \simeq 93$~MeV, $N_\pi = 3$ is the number of intermediate pions, and $I(x,y)$ is the imaginary part of the loop function. The absorptive part of the amplitude gives an estimate of the minimum value for this rate. We find
\begin{align}
    \mathcal{B}(K_{S,L}\to aa) &> \left(\frac{m_a^2 N_\pi}{32 \pi F_\pi^2}\right)^2\mathcal{B}(K_{S,L}\to \pi^0\pi^0)
    \\\nonumber
    &\qquad\times\lambda^{1/2}(1,r_a^2,r_a^2)\lambda^{1/2}(1,r_\pi^2,r_\pi^2)
   \\\nonumber
    &\simeq
    \left\{\begin{matrix} 
    2.6\times10^{-7} &\text{ for }K_S,
    \\
    7.2\times10^{-10} &\text{ for }K_L.
    \end{matrix}\right.
\end{align}
While we do not calculate it here, we note that the dispersive part of the amplitude may actually give a comparable or larger contribution, with some pieces being enhanced by a formally large logarithmic factor, $\sim\log(4\pi F_\pi/m_K)$.

The $K_S$ rates drastically exceed the SM expectation for $K_S\to2(e^+e^-)$, but no searches for this decay, with or without a bump hunt, have been performed. As discussed in \refsec{sec:decays}, future efforts at LHCb may be capable of probing this decay mode~\cite{Dettori:2019oak}. With ${\cal O}(10^{-7}-10^{-6})$ branching ratio predicted in this model, even the KLOE experiment may have sensitivity to $K_S\to 2a\to4l$.
For the $K_L$ decays, the current distributions available in the $K_L\to 2(e^+e^-)$ measurement by KTEV are not sensitive to such small branching ratios.  This could change with additional information on invariant mass distributions and by searching for a double coincidence in $m_{ee}=m_{ee}^\prime = m_a$.

\subsubsection{$K\to\pi\pi a a$}

Finally, we note that the double axion production can also happen with two real pions, where again the rate can estimated by the use of \refeq{eq:aapipivertex} and existing knowledge of $\mathcal{M}(K\to\pi\pi)$. We include double axion emission from the pion lines, which constructively interfere, as well as from the initial state kaon. The latter destructively interferes with the rest but is sub-dominant. The relevant vertex is simply $\mathcal{M}_{K-K-a-a} = (m_a^2/F_\pi^2)/3$. Using the phase space decomposition in Ref.~\cite{Cappiello:2011qc}, we compute the rates for $a(17)$ to find
\begin{align}  \mathcal{B}(K_{S}\to \pi^+\pi^- a a) &\simeq 5\times10^{-9},
\\
\mathcal{B}(K_L\to \pi^+\pi^- a a) &\simeq 1\times10^{-11},
\\
\mathcal{B}(K^+\to \pi^+\pi^0 a a) &\simeq 1\times10^{-9}.
\end{align}
Out of the channels above, $K_S$ decays are the only ones that do not compete with large backgrounds from $K\to\pi\pi \pi^0_{\rm DD}$ decays in the SM. This six-track signature appears at rates much below $K\to\pi aa$, so we do not discuss it further.

\subsection{Triple axion production}

The non-linear structure of the QCD axion interactions does not stop at $2a-2\pi$ vertex, and can induce multiple production of $a$. Indeed, since $f_a$ is not particularly large, higher-order terms in $a$ can also be important. We notice that there is sizable triple production of axions in $\pi^0,\eta,K_L\to 3a$. The decays of $\eta$ and $K_L$ may depend on details of ChPT and/or interference of several contributions. However, the decays of $\pi^0\to 3a\to 6l$ can be predicted with no free parameters, other than the mass of an axion. 
    
The $\pi^0a^3$ interaction survives even in the extreme ``pio-phobic" case, $m_u=m_d/2$ and $Q_d = Q_u/2 = 1$, and can be easily computed from the axion-dependent chiral Lagrangian. Expanding it to an appropriate order, we get 
\begin{equation}
    {\cal L}_a = ... +\frac16 \frac{m_a^2}{F_\pi f_a} a^3\pi^0+ 
        (a^3\eta_8, a^3\eta_s~{\rm terms}).
\end{equation}
This immediately leads to the following prediction: 
\begin{equation}
    \left.{\cal B}(\pi^0\to 3a\to 3(e^+e^-))\right|_{m_a=17\,
    {\rm MeV}} = 1.0\times 10^{-3}. 
\end{equation}
It would be appropriate to say that this is a gigantic rate, and it would be indeed the third largest branching after $\gamma\gamma$(0.99) and $\gamma e^+e^-$(0.01), exceeding the SM double-Dalitz decay by a factor of 30. We believe that such a large rate should have been noticed, {\em e.g.} in the studies of 
$\pi^0\to  2(e^+e^-)$ via capture of $\pi^-$ 
\cite{Samios:1962zza}. (In that work, double-Dalitz and single-Dalitz decays were observed by human examination of photographs from a bubble chamber, and missing very frequent 6-track decays seems implausible.)

On the other hand, we could not find any immediate exclusion of such a large exotic rate based on existing searches. When the rate of $\pi^0$ decay to six leptons is that high, it would most certainly create new sources of background to many other rare radiative decay searches, of $K$-mesons in particular. What seems not controversial, however, is that dedicated searches in a large and controlled sample of $\pi^0$, either from $K$ decays or $\pi^-p$ conversion, should unambiguously settle this issue. Another promising way to study $3a$ decay modes of $\pi^0$ would be via the production of vector resonances at $e^+e^-$ machines. For example, the branching of $\phi \to \pi^+\pi^-\pi^0 \to \pi^+\pi^-3(e^+e^-)$ is expected to be $1.5\times 10^{-4}$ for $a(17)$, which will result in a large number of events at KLOE. Other examples may include $\tau^+\tau^-$ pairs at $B$-factories, which produce $\pi^0$ in 25\% of events.

Finally, although less controlled, neutrino experiments also offer large samples of $\pi^0$ events, as for example MiniBooNE~\cite{AguilarArevalo:2009ww}, where we estimate a total number of $\sim1.7\times10^{5}$ neutral pions were produced in neutrino scattering given the latest number of protons on target~\cite{MiniBooNE:2020pnu}. Other detectors with improved particle identification may be more suitable for dedicated searches, such as NOMAD~\cite{Kullenberg:2009pu} and $\mu$BooNE~\cite{Adams:2018sgn}, where we expect a total of $\mathcal{O}(3\times 10^4)$ neutrino-induced $\pi^0$'s in their full exposure.

\subsection{Pion capture $\pi p \to aa n$}

We now turn our attention to an old technique to study meson properties: pion capture. The possibility of a single axion production, along with the single production of other exotic particles, has been addressed in a recent publication~\cite{Chen:2019ivz}.
In the pio-phobic model of Ref.~\cite{Alves:2020xhf}, it is conceivable that such amplitudes are tuned to zero using mixing angles between axions and pions. Therefore, we concentrate on double axion production that must exist at an appreciable level, again due to $a-a-\pi-\pi$ vertex \refeq{eq:coupling_pi2a2}. Unlike the case of $\eta$, $\eta'$ and $K$ meson decays, where the mass scale of decaying particles is not too far from where one expects a breakdown of ChPT, the threshold reaction of $\pi^-$ on protons must be fairly reliably described by the leading order ChPT. 

\begin{figure}[t]
    \centering
    \includegraphics[width=0.45\textwidth]{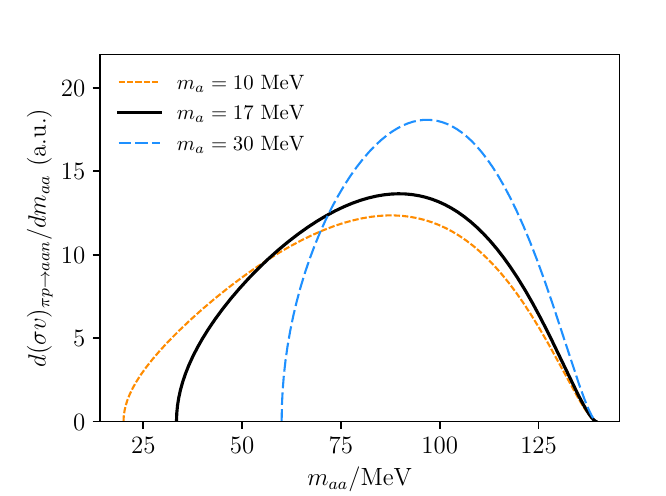}
    \caption{The differential pion capture cross section as a function of $m_{aa}$ for three QCD axion masses. Clearly, most events peak away from the pion mass, where pion double-Dalitz decays become a background.}
    \label{fig:picapture_maa}
\end{figure}

The coupling of pions to nucleons is well-known in ChPT, and we can compute the rate for pion capture from \refeq{eq:coupling_pi2a2} and 
\begin{equation}
    \mathcal{L} \supset -\frac{g_A}{\sqrt{2} F_\pi} \overline{n} \gamma^\mu \gamma^5 p (\partial_\mu \pi^-) + \text{h.c.}, 
\end{equation}
where $g_A$ is the nucleon axial-vector coupling. Neglecting form factors, nucleon mass differences and magnetic moments, we find
\begin{widetext}
\begin{align}
    \frac{\dd (\sigma v)_{\pi^- p^+ \to a a n}}{\dd m_{aa}^2 \dd m_{a n}^2} = 
    \frac{ g_A^2 m_a^4}{512 \pi^3 F_\pi^6}
    \frac{ M^2(m_\pi^2 - m_{aa}^2) }{m_\pi (M+m_\pi)(m_{aa}^2 M -m_\pi^2 (2 M + m_\pi))^2 },
\end{align}
\end{widetext}
where $M$ is the nucleon mass. For direct comparison with experiment, one can compute the branching ratio
\begin{align}\label{eq:capturerate}
    \mathcal{B}(\pi^- p^+ \to a a n) &= \frac{(\sigma v)_{\pi p \to a a n}}{ (\sigma v)_{\pi p\to\pi^0 n} + (\sigma v)_{\pi p \to \gamma n}} 
    \\\nonumber
    &\simeq 6.2 \times 10^{-7},
\end{align}
indicating that the signal rate is far below that of pion production. In fact, the relevant rate to compare \refeq{eq:capturerate} to is double radiative pion capture, $\pi^- p^+ \to \gamma \gamma n$. It was most recently measured at TRIUMF~\cite{Tripathi:2007gz} with
\begin{equation}
\mathcal{B}(\pi^- p^+ \to \gamma \gamma n) = (3.02 \pm
0.27(\text{stat.}) \pm 0.31(\text{sys.}))\times 10^{-5}.
\end{equation}
Since the two gammas are detected after converting, this channel is effectively a 4 lepton final state measurement. Nevertheless, the current measurement is not sensitive to branching ratios below $10^{-6}$, but a dedicated search is well within experimental reach. Other sources of backgrounds are double-Dalitz decays, at a rate of $\mathcal{B}(\pi^- p^+ \to \pi^0_{\rm DD} n \to 2(e^+e^-) n) \simeq 2\times 10^{-5}$. (The first observation of the double-Dalitz decay of $\pi^0$ was in fact performed using the $\pi^-$ capture on protons~\cite{Samios:1962zza}.) Nevertheless, the invariant mass of the four-lepton system shown in \reffig{fig:picapture_maa} peaks at values much smaller than that of the neutral pion, and would allow to largely reject the pion-production-induced backgrounds. Moreover, the $\pi^0$-induced background can be significantly reduced by using the capture on deuterium. One should expect that the $2a$-induced $4l$ capture rate in Deuterium, 
${\rm D} + \pi^- \to nn+ aa\to nn+2(e^+e^-)$ to dominate any SM sources of $4l$, and therefore it also provides a reliable way of resolutely testing the $a(17)$ QCD axion model. Should such a search be pursued, it can also refute (or discover) $\pi^-p\to n\pi^0\to n3a\to n3(e^+e^-)$ predicted for $a(17)$ to have ${\cal B} =0.6\times 10^{-3}$. 

\subsection{Radiative pion decays}

In this sub-section, we evaluate possible sensitivity that can be achieved via rare pion decays. Radiative pion decay~\cite{Egli:1989vu}, $\pi^+\to \nu e^+ a\to \nu e^+ e^-$ provides an important constraint on the emission of light particles (see {\em e.g.} Ref.~\cite{Egli:1989vu,Dror:2017nsg}). The $\pi^0-a$ mixing leads to the beta-decay type transition, $\pi^+ \to \nu e^+ (\pi^0)^* \to \nu e^+ a$, and was used to constrain $\theta_{\pi^0a}$~\cite{Alves:2020xhf}. But even if the mixing angle vanishes, the $a$ emission from the electron line is possible. For $a(17)$, at $\theta_{\pi^0a} =0 $, the predicted branching ratio is ${\cal B}(\pi^+\to \nu e^+ a) = 2\times 10^{-9}Q_e^2$, where $Q_e$ is the Peccei-Quinn charge of electron. Results of~\cite{Egli:1989vu,Dror:2017nsg}) imply $Q_e < 0.5$, that together with NA64 constraints~\cite{NA64:2020xxh} leave only a small part of the parameter space viable for $a\to e^+e^-$-based phenomenological explanations of 17\,MeV anomalous signals.

\section{Conclusions}

Over the years, rare kaon decays have been instrumental in learning about the properties and nature of weak interactions, $P$ and $CP$ violation, and one of the main drivers in establishing the Standard Model. In parallel, the study of K-mesons and their decay products has also played a major role in the search for new physics, providing some of the stringiest limits on the existence of light new particles. In this paper, we have argued that these studies are far from being complete, as a number of very promising decay modes have escaped experimental attention so far. In particular, the production of new visible resonances in multiplicities of $n>1$ could lead to multi-leptons final states that have never been searched for in kaon decays. We provide a simple example of a dark scalar in $K\to \pi \hdark$ that fragments into pairs of dark states, such as dark photons, each of which decays to an electron-positron pair. The rates of $K$ decays to $\pi+2(e^+e^-)$ due to such DS cascades can significantly exceed corresponding SM rates, and may even dominate over the $\pi^0$ Dalitz-induced backgrounds. Such multi-lepton rates do not necessarily compete with existing searches for single $e^+e^-$ final states or for missing mass in, for example, $K\to \pi \slashed{E}$, and can probe unexplored parameter space. The ``bump-hunt" on top of the smooth distribution of lepton pairs over the invariant mass will provide further discriminating power for separating DS signatures over the background, and in the case of discovery, revealing the mass of either one or several dark states. We have also pointed out new decay channels that can probe short-lived heavy neutral leptons, in their non-minimal modifications. The production of $N$ via the neutrino portal will involve final states with at least one neutrino, and constitutes a natural extension of peak searches in $K^+\to\ell^+ N$, where the additional multi-leptons would be experimentally vetoed.

We also made some predictions of rare kaon decays within a recently proposed model of a MeV-scale QCD axion~\cite{Alves:2017avw,Alves:2020xhf}. While it is true that the single axion production in this type of model suffers from QCD and model-building uncertainties, we argue that $aa$ and $aaa$ production in the final states of meson decays are more robust. While the limit of $Q_{d}=Q_u/2 = 1$ and $m_u=m_d/2$ ensures that the axion mixing with pions vanishes, other $a-\pi^0$ interaction terms of higher-order in the axion field do not, all of which are a direct consequence of QCD. For the $a(17)$, motivated by recent anomalies in nuclear de-excitation, the quadratic ($a-a-\pi-\pi$) and cubic ($a^3\pi^0$) vertices are not particularly small, $(m_a/F_\pi)^2 = 0.033$ and $m_a^2/(F_\pi f_a) = 0.0030$, respectively. As a consequence of the quadratic vertex, one should expect $K\to \pi+2(e^+e^-)$ branching ratios {\em to exceed} ${\cal O}(10^{-5})$ levels, which can be easily probed by using existing data sets.
Even if one assumes that the $\Delta S=1$ $K-\pi$ mixing is very small,  the well-known $K\to 3\pi$ amplitude will generate $K\to \pi aa$ at one-loop level at a measurable rate. 

Similarly, we found large $2a$ production, with branching ratios of ${\cal O}(10^{-6})$, in the negative pion capture on protons and deuterium, which can be probed at light meson facilities. Finally, we have calculated the triple emission of the pio-phobic $a(17)$ in $\pi^0$ decays, finding an extremely large result,
${\cal B}(\pi^0\to 3(e^+e^-)) \simeq 10^{-3}$. If this model was indeed realized in nature, the axion-induced triple-Dalitz rate would exceed double-Dalitz decay of $\pi^0$ by a factor of $\sim 30$ at $m_a=17$\,MeV. Thus, rare decays of $\pi^0$ would also prove key in excluding the MeV-scale axion models. The test of such an unusual $\pi^0$ decay can be made with a dedicated $\pi^--p$ capture experiment, as well as by sourcing the $\pi^0$ via large $K$-decay modes: $K^+\to \pi^+\pi^0,\,K_L\to 3\pi$.

It is important to emphasize that not every model aiming at explaining ``17 MeV anomaly" in nuclear decays is probed by the decays discussed in our paper. Specifically, we have pointed signatures of the  highly non-linear MeV-scale axion model. In other incarnations of $X(17)$ that involve vector particles~\cite{Feng:2016jff,Feng:2016ysn} or axion-like particles with much larger $f_a$~\cite{Ellwanger:2016wfe}, the multiple emission of $X$ can be suppressed. However, the kaon physics
(and flavor physics in general) would still provide important constraints on these models via, for example, $K_L\to \gamma X\to \gamma (e^+e^-)$, $K_L\to \pi\pi X \to \pi\pi(e^+e^-) $ and $\Sigma^+\to pX\to p(e^+e^-)$. The constraining power of these measurements, in connection with a variety of $X(17)$ models, will be addressed in a subsequent publication.

We close by stressing that several of the decay rates we discuss typically exceed the Standard Model as well as the intrinsic background rates, some of which have never been searched for. While dedicated searches for multiple visible resonances would constitute experimentally intensive work, the sheer number of unexplored combinations of $K\to n(e^+e^-) + m(\pi^0) + (\ell^+\nu)$ modes, with $n\geq2$ and $m=\{0,1,2\}$, warrants a broader and less tailored approach. Indeed, fully inclusive measurements of multi-lepton final states could already provide new sensitivity even with minimal understanding of the backgrounds and without detailed study of the kinematics. Once such an exercise is carried out, more exclusive searches can be performed, including multi-dimensional ``bump hunts" in the various combinations of lepton masses. Having identified clear targets for the multi-lepton branching ratios in axion and dark sector models, we hope that future work will help elucidate more experimental aspects of multi-leptons at kaon factories. Understanding trigger and detector acceptances to multi-leptons will be crucial to derive the optimal experimental sensitivities to the channels in \reftab{tab:BRstable}.

\begin{acknowledgments}
We would like to thank Dr.~Robert Tschirhart for correspondence on the KTEV experiment and Dr.~Daniele Alves for discussions about MeV-scale QCD axions.
MP is supported in part by U.S. Department of Energy (Grant No. desc0011842). This research was supported in part by Perimeter Institute for Theoretical Physics. Research at Perimeter Institute is supported by the Government of Canada through the Department of Innovation, Science and Economic Development and by the Province of Ontario through the Ministry of Research, Innovation and Science.
\end{acknowledgments}
%
\appendix

\section{Decay rates for $K \to\Adark\Adark$}
\label{app:calculation_rate}
\begin{figure*}[t]
    \centering
    \includegraphics[width=0.45\textwidth]{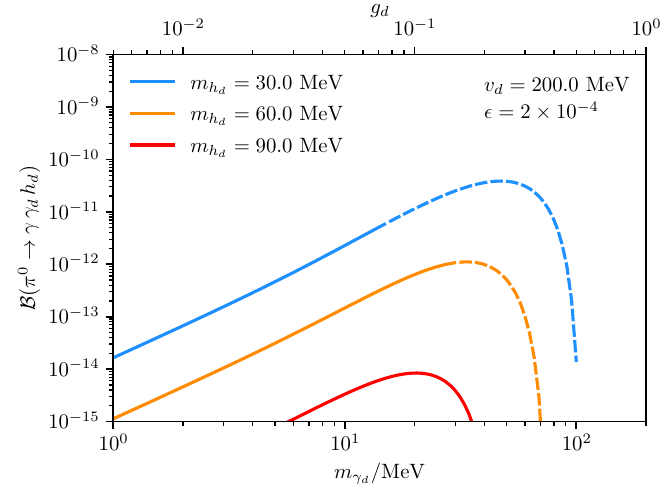}
    \includegraphics[width=0.45\textwidth]{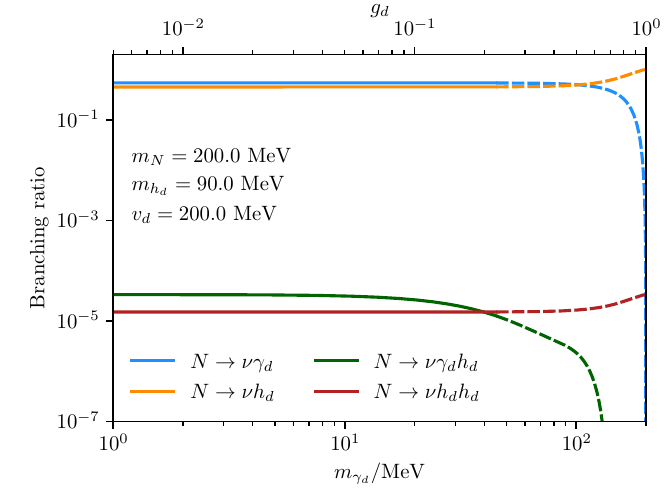}
    \caption{On the left we show the typical branching ratios for $\pi^0\to\gamma\Adark\hdark$, and on the right plot we show the dark sector branching ratios for $N$. Solid (dashed) lines indicate that $\hdark \to \Adark\Adark$ decays are kinematically allowed (forbidden). In varying the dark photon mass, we also vary the dark coupling, shown as a secondary axis in both figures. \label{fig:6lepton_BR}}
\end{figure*}
\subsection{Minimal higgsed $U(1)_d$}

After integrating out the Higgs, the dark scalar couples to the SM flavor-changing neutral-current (FCNC),
\begin{equation}
    \mathcal{H}_{\rm FCNC} \supset g_{sd} m_s \overline{s_R} d_L \times(\lambda_d v_d \hdark) + \text{h.c.},
\end{equation}
where $g_{sd}$ was defined in \refeq{eq:gsd}. For $K_{S,L}\to\Adark\Adark$ decays, we are interested in the $K_{S,L}$ transitions to vacuum. Neglecting small $CP$ violating terms, we write
\begin{equation}
    \bra{0} g_{sd} m_s \overline{s_R} d_L + \text{h.c.} \ket{K_L} = \Im(g_{sd}) F_K m_K^2,
\end{equation}
where $F_K\simeq 117$~MeV. The resulting amplitude can then be related to $h\to\Adark\Adark$, and since both are independent of the phase space, we quote only the total decay rates as
\begin{align}
    \Gamma_{h\to \Adark \Adark} &= \frac{\lambda_d^2}{64\pi} \frac{m_h^3v^2}{(m_h^2 - m_{\hdark}^2)^2} f(r_h),
    \\
    \Gamma_{K_L\to \Adark \Adark} &= \frac{\left( \lambda_d \Im(g_{sd})\right)^2}{64\pi} \frac{m_K^7F_K^2}{(m_K^2 - m_{\hdark}^2)^2} f(r_K),
    \\
    \mathcal{B}({K_L\to \Adark \Adark}) &\simeq 5 \times 10^{-18} \times  \left(\frac{\mathcal{B}({h\to \Adark \Adark})}{10\%}\right)
\end{align}
where $f(r_P)$ is defined below \refeq{eq:hdarkdecays}. We neglect final state masses when proving our numerical estimate. 

\subsection{$U(1)_d$ plus scalar singlet}
The FCNC hamiltonian in this case depends on the super-renormalizable portal coupling between the singlet and Higgs sectors,
\begin{equation}
    \mathcal{H}_{\rm FCNC} \supset g_{sd} \, m_s \overline{s_R} d_L \times(A_H S) + \text{h.c.}
\end{equation}
The resulting amplitude is simply,
\begin{align}
    \mathcal{M}_{K_L\to \Adark\Adark} &= \Im(g_{sd}) F_K m_K^2 \times
    g_d^2 v_d      \\\nonumber
&\qquad\times \frac{v A_H}{m_K^2 - m_{S}^2 + i \Gamma_S m_S}\times\epsilon^{\mu*}(k_1) \epsilon^*_{\mu}(k_2),
\end{align}
and can then be related to $h\to\Adark\Adark$. Since both amplitudes are independent of the phase space variables, we simply quote the total decay rates,
\begin{align}
    \Gamma_{h\to \Adark \Adark} &= \frac{1}{64\pi} \frac{m_h^3v^2}{(m_h^2 - m_{S}^2)^2} \frac{A_H^2 A_\phi^2 }{(m_{\hdark}^2-m_S^2)^2}  f(r_h),
    \\
    \Gamma_{K_L\to \Adark \Adark} &= \frac{\Im(g_{sd})^2}{64\pi} \frac{A_H^2 A_\phi^2 }{(m_{\hdark}^2-m_S^2)^2}
    \\ \nonumber
&\quad\quad\times \frac{m_K^7 F_K^2 v^2}{(m_K^2 - m_{S}^2)^2+m_S^2\Gamma_S^2} f(r_K).
\end{align}

\section{Dark higgstrahlung decays}\label{app:higgstrahlung}

Below the pion mass, one has the advantage that the $\pi^0$ decays electromagnetically and dark photon production can proceed via kinetic mixing, being suppressed only by $\epsilon^2$ rather than $\alpha \epsilon^2$. The two-body decay $\pi^0\to(\Adark\to e^+e^-) \gamma$ has been searched for at SINDRUM~\cite{MeijerDrees:1992kd}, WASA-at-COSY~\cite{Adlarson:2013eza}, and NA48/2~\cite{Batley:2015lha}, all sensitive to $\epsilon$ of $\mathcal{O}(10^{-3})$. If the dark scalar is sufficiently light, another decay channels opens up, namely 
\begin{equation}
\pi^0 \to \gamma \Adark \hdark \to \gamma 3(e^+e^-).    
\end{equation}
The branching ratio was computed in Ref.~\cite{Batell:2009di} and is rather small, reaching values of $\mathcal{O}(10^{-11})\times(\epsilon/2\times10^{-4})^2$ for the lightest dark photons and decreasing rapidly for larger $m_{\hdark}$.

On top of the single dark photon decay of the HNLs, we also considered dark higgstrahlung and double dark Higgs emission. These rates, however, are small, just below $\mathcal{O}(10^{-5})$. The decay chains would appear as
\begin{align}
    K^+&\to\ell^+ (N \to \nu \Adark \hdark \to \nu (e^+e^-) 2\Adark \to \nu 3(e^+e^-)),
    \\
    K^+&\to\ell^+ (N \to \nu \hdark \hdark \to \nu 2\Adark 2\Adark \to \nu 4(e^+e^-)),
\end{align}
when produced through the neutrino portal. A comparison with the dominant branching ratios is provided in \reffig{fig:6lepton_BR}. Picking the largest allowed mixing angles, would still render the overall rate for these processes to be below the $10^{-10}$ level.

\bibliographystyle{apsrev4-1}
\bibliography{main}{}


\end{document}